# THE SECOND HARMONIC GENERATION IN THE SYMMETRIC WELL CONTAINING A RECTANGULAR BARRIER

D.M.Sedrakian [1], A.Zh.Khachatrian [2], G.M.Andresyan [3], V.D.Badalyan [1]

[1] Yerevan state university 375025, Yerevan, Armenia, Alex Manukyan 1

[2] The state engineering university of Armenia, 375046 Yerevan, Armenia, Teryan 109

3 Yerevan state university of architecture, 375046 Yerevan, Armenia Teryan 106

The problem of determination of the maximum of second harmonic generation in the potential well containing a rectangular barrier is considered. It is shown that, in general, the problem of finding the ensemble of structures with equidistant first three levels has two types of solutions. For the first type the second and third energy levels are located above a rectangular barrier, and for the second type the third level is located above the barrier only. It is also shown, that generation corresponding to the second type of solution always is less than generation for the first one. Taking into account the effective mass changes the problem of finding the generation maximum for a finite depth well is exactly solved.

## 1. Introduction

Thanks to the recent progresses in epitaxy technology, the fabrication of ultra-thin layered systems from semiconductor materials with various values of width of the forbidden bands and electronic affinity has become possible. To get the required characteristics of various one-electronic devices, heterostructures from two wells of *GaAs* dividing by a potential barrier of *AlAs* is usually used [1-3]. It is well known that in symmetric wells the selection rule takes place, and overlapping integrals for states belonging to different subband of quantized spectrum are equal to zero or unit. However, in many interesting cases the quantum wells with a complex structure or simple wells made asymmetric by application of electric field are examined. In these cases the selection rule is broken, this leads to changes the radiation and absorption conditions for light in such systems.

As it is known in the semiconductor size-quantized structures, the dipole matrix elements have the same order of magnitude as the quantum well width (a few nanometers). The last leads to extremely large nonlinear susceptibility for such structures in comparison with molecular and ionic systems, in which the dipole matrix elements are about few picometers. As a result of the subband and band - to - band transitions, the problem of second harmonic generation (SHG) for an electromagnetic radiation in the infra-red region of wavelengths, recently the study of every possible asymmetric low-dimensional structures having equidistantly located energy levels has been the subject of considerable attention [4-12].

The intersubband SHG coefficient is given by [13]:

$$\chi^{(2)}_{2\omega} = \frac{q^3}{\varepsilon_0} \sum_{ij} \frac{1}{2\hbar\omega + E_{ij} - i\Gamma_{ij}} \sum_k \mu_{ij}\mu_{jk}\mu_{ki} \left[ \frac{\rho_i - \rho_k}{\hbar\omega + E_{ik} - i\Gamma_{ik}} - \frac{\rho_k - \rho_j}{\hbar\omega + E_{kj} - i\Gamma_{jk}} \right], \quad (1)$$

where $E_{lm} = E_l - E_m$ is the transition energy between $l$ and $m$ subbands, $\mu_{lm} = \langle l | z | m \rangle$ are dipole matrix elements, $\hbar\omega$ - the photon energy, $\rho_l$ - is the surface charge density of $l$ subband and $\Gamma_{lm}$ - accounts for inhomogeneous broadening of $l$ to $m$ transition. Most studies of the SHG focus on the so-called double resonance regime, when in the three-level system the generation coefficient is maximal. In this case the system's levels are located equidistantly and the photon energy $\hbar\omega$ is equal to the energy difference $1 \to 2$ and $2 \to 3$ transitions. Hence, at the double resonance regime Eq.(1) can be written as

$$\chi^{(2)}_{2\omega} = \frac{q^3 \rho_1}{\varepsilon_0} \frac{\mu_{12}\mu_{31}\mu_{23}}{(2\hbar\omega - E_{13} - i\Gamma)(\hbar\omega - E_{12} - i\Gamma)}, \quad (2)$$

where $\Gamma_{12} = \Gamma_{13} = \Gamma$ is taken. According to Eq.(2), the SHG can be observed for asymmetric potential only, since in that case all dipole matrix elements are differ from zero and therefore $\mu_{12}\mu_{31}\mu_{23} \neq 0$.

For the given configuration of confinement potential at the double resonance the SHG can be maximized by means of variation of structure parameters. To do this it is necessary except the determination of the ensemble of structures with given value of levels equidistant to determine the such structure, for which the product of dipole matrix elements is maximum. In the general case, the determination of optimal confinement potential is a very difficult nonlinear variation problem. Moreover, both the structural and material parameters of the optimal potential depend on the radiation frequency.

In papers [13, 14] the problem of determination of the SHG maximum was considered for an infinite deep well containing a rectangular barrier directly adjoining to the wall of the well. Regarding to the above mentioned papers for the given type of asymmetric potential the procedure of determination of optimal structure can be done at once for all wavelengths. The last has allowed definition of the structure constructing on the basis of materials $Al_nGa_{1-n}As$ with rather large coefficient of SHG in the infra-red region of frequencies and to experimentally observe the generation. At the same time the authors of papers [13, 14] did not receive optimal parameters for finite depth well. As our consideration shows for the well chosen in papers [13, 14] for experimental observation of SHG the condition of equidistant energy levels does not take place. Moreover, for exact realization of the structure one has to take into account the effective mass changes in the different areas of the heterostructure also.

This paper is devoted to the problem of determination of optimal parameters for the structure, presenting the finite depth well containing a rectangular barrier adjoining to the one wall of the well, which give the maximum value for the SHG coefficient taking into account the effective mass changes.

## 2. The maximum of the SHG coefficient for infinite deep quantum well

Let us consider the problem of finding the possible values of parameters for the infinite deep well with adjoining to the one wall rectangular barrier, for the which the first three energy levels are equidistantly located from each other. The potential has the form of:

$$V(x) = \begin{cases} \infty, & x < 0 \\ 0, & 0 < x < d \\ U, & d < x < L \\ \infty, & x > L \end{cases} \quad (1)$$

The given potential has the three independent parameters: $U$ - value of rectangular barrier potential, $b = L - d$ is the thickness of rectangular barrier, $L$ - the width of infinite deep well.

To solve the formulated problem we shall consider the equation determining the electron energy spectrum for potential (1). Without taking into account the effective mass changes this equation looks like [15]

$$f(E) \equiv \sin\{kL\}(\operatorname{Re}\beta - \operatorname{Re}\alpha) - \cos\{kL\}(\operatorname{Im}\alpha - \operatorname{Im}\beta) = 0, \quad (2)$$

where

$$\alpha = \exp\{-ikb\}\left[\cos\{qb\} + i\frac{q^2 + k^2}{2kq}\sin\{qb\}\right], \qquad (3)$$

$$\beta = i\exp\{-i2(L - d/2)k\}\frac{q^2 - k^2}{2kq}\sin\{qb\}. \qquad (4)$$

In Eq. (2) - (4) the following designations are made: $k^2 = 2mE/\hbar^2$ and $q^2 = 2m(E - U)/\hbar^2$, where $E$ is the electron energy and $m$ is its effective mass. Let us denote thought $E_n$ the roots of Eq.2, then one can write down it in the following form:

$$f(E_n, U, L, b) = 0.$$

This equation defines $E_n$ at given values of three parameters $U, L, b$. According to Eq. (3), (4), it can be considered as a connection between three dimensionless parameters $e_n = \frac{2m}{\hbar^2}E_n b^2$, $u = \frac{2m}{\hbar^2}Ub^2$ and $x = b/L$:

$$f(e_n, u, x) = 0. \qquad (5)$$

Let us now require the lower three levels to be located equidistant from each other: $e_3 - e_2 = e_2 - e_1 = \Delta$, where $\Delta$ - distance between levels. According to (5) this requirement is equivalent to the following three equations:

$$f(e_1, u, x) = 0, \qquad (6)$$

$$f(e_1 + \Delta, u, x) = 0, \qquad (7)$$

$$f(e_1 + 2\Delta, u, x) = 0. \qquad (8)$$

At any fixed value of $x$ Eq. (6) - (8) can be considered as the set from three transcendental equations for three unknown parameters $e_1$, $\Delta$ and $u$. As consideration shows the set (6) - (8) has solutions for values $x$ more than $x = 0.4069$ only. For each value of $x$ there are two collections of the quantities $\{e_1, \Delta, u\}$ satisfying the set of equations (6) - (8). So, for example, at $x = 0.41$ we have $\{3.0729, 7.9236, 8.9329\}$ and $\{3.2696, 8.8434, 12.2986\}$. Each of collection of the quantities $\{e_1, \Delta, u\}$ represents a separate type of solution for the equations (6) - (8). In the figure 1 for the first type of solution we plot the dependencies of quantities $e_1, e_2, e_3, u$ and $\Delta$ from the variable $x$.

As one can see from the figure for any value $x$ the positions of the second and third levels are in over-barrier region of energies. It should be noted, that according to the right side of figure 1, there are two potential wells for which the distance between the levels is equal to the value of potential of rectangular barriers and there is one potential well for which distance between the levels is equal to the value of energy of the first level.

In figures 2 for the second type of solution we plot the dependencies of quantities $e_1, e_2, e_3, u$ and $\Delta$ from the variable $x$. As seen from the figure for at any value of $x$ the first and second levels always are in under barrier region of energies.

Taking the above-mentioned results as a base, we have calculated the product of normalized dipole matrix elements of transitions $\chi = |\mu_{12}\mu_{23}\mu_{31}|/Ld_\nu^2$ ($h\nu$ is photon energy of a and $d_\nu = \hbar\pi/\sqrt{2mh\nu}$ - the width infinite deep well with energy of the first level equals to photon energy) depending on dimensionless parameter $x$.

As seen from the figure 3a, the coefficient of generation gets its maximum value at x = 0.5685. According to the result of paper [13] $\chi$ is a maximum at x = 0.5677, that means, that the result obtained here coincides with result of paper [13]. As is visible from the figure 3в for the second type of solution the maximal generation takes place at the smallest value of parameter $x$. Essentially here we have the intensity of generation corresponding to the second type of solution always remains smaller then the generation corresponding to the first type for any $x$. The last allows the problem to be completely solved, being limited by the first type of solution only.

According to the above-mentioned, independent from the photon frequency the SHG becomes the maximum when dimensionless parameters $x$, $\Delta$, $u$ take the certain values, namely

$$x_{max} = \frac{d}{L} = 0.5685 \quad \Delta_{max} = \frac{2m}{\hbar^2}h\nu d^2 = 14.58 \quad u_{max} = \frac{2m}{\hbar^2}Ud^2 = 15.82. \quad (9)$$

At the given photon energy $h\nu$, according to (9), the parameters of the potential are defined as follows:

$$d = 2.7\frac{\hbar}{\sqrt{mh\nu}} \quad L = 4.75\frac{\hbar}{\sqrt{mh\nu}} \quad U = 1.08h\nu. \quad (10)$$

As seen from Eq.(10) the increase of photon energy leads to the decrease of the barrier and well widths while the height of the barrier increases. At electron energy $h\nu = 117.16\,meV$ [14] from Eq.(10) for parameters of the well we have $d = 8.48\,nm$, $L = 13.94$ and $U = 126.5\,meV$. As we shall see below, the account of finiteness of the well brings to the significant corrections in the values of the well parameters.

## 2. Generation in the finite depth well

Above we have considered the problem of finding the maximum of SHG for infinite deep well containing inside a rectangular barrier, without taking into account the changes of effective mass. We have actually repeated calculations of work [14] by using the method suggested in the paper [15] for demonstration of its accuracy and efficiency.

In the given section we will consider the same problem, but for a finite depth well taking into account the changes of electron effective mass in difference layers of heterostructure. Dependencies of an electron potential energy and its effective mass from the coordinate have the form:

$$U(x), M(x) = \begin{cases} V, M, & x < 0, \\ U, m, & 0 < x < d, \\ 0, m_0, & d < x < L, \\ V, M, & x > L. \end{cases} \quad (11)$$

As can be seem from Eq.(11) the considered potential has four independent parameters $V, U, d, L$ in contrast to the potential (3) which had three independent parameters. This essentially makes the solution more difficult compared with the solution considered in the Sec.2. In particular, here for the each value of photon energy the optimization of SHG should be done separately. For potential (11) the energy spectrum is determined from the following equation [15]:

$$tg\{kd\} = \frac{2m_0 M \chi k \,\mathrm{Re}(1/t) + (m_0^2 \chi^2 - M^2 k_0^2)\,\mathrm{Im}(1/t) - (m_0^2 \chi^2 + M^2 k^2)\,\mathrm{Im}(r/t)}{2m_0 M \chi k \,\mathrm{Im}(1/t) - (m_0^2 \chi^2 - M^2 k^2)\,\mathrm{Re}(1/t) + (m_0^2 \chi^2 + M^2 k^2)\,\mathrm{Re}(r/t)}, \quad (12)$$

where the following designations are entered

$$\frac{1}{t} = \exp\{ikd\}\left[\cos\{qd\} - i\frac{m_0^2 q^2 + m^2 k^2}{2m_0 mqk_0}\sin\{qd\}\right], \quad (13)$$

$$\frac{r}{t} = \frac{i(m_0^2 q^2 - m^2 k^2)}{2m_0 mqk}\exp\{ikd\}\sin\{qd\}, \quad (14)$$

where

$$\chi = \sqrt{\frac{2M}{\hbar^2}(V-E)} \quad q = \sqrt{\frac{2m}{\hbar^2}(E-U)} \quad k = \sqrt{\frac{2m_0}{\hbar^2}E}. \quad (15)$$

Note, that in Eq. (13), (14) $r$ and $t$ are an electron reflection and transmission amplitudes for the rectangular barrier with the center at the point $x = d/2$. Further, we shall examine the structure prepared from the concrete material $Al_n Ga_{1-n} As$ ($0 \leq n \leq 0.4$) for which of an electron potential energy and effective mass in the conductivity band have the form of [14,16, 17]:

$$U(n) = 0.6(1.36n + 0.22n^2)eV \quad m(n) = m_e(0.067 + 0.083n), \quad (16)$$

where $m_e$ - free electron mass. Note, that in Eq.(16) the energy corresponding to the conductivity band bottom of $GaAs$ is zero. Further, we shall put that well walls are prepared from the material $Al_{0.4}Ga_{0.6}As$ which corresponds to well depth about $347\,meV$.

It is important to note that as in the case of infinite deep well the finite depth well has two types of solutions also. The first is that when one level is in underbarrier region only and the second one is that when in underbarrier region there are two levels. Due to that for infinite deep well the SHG coefficient corresponding to second type of solution is always less than the SHG coefficient corresponding to the first type (see above) at any values of problem parameters, for the potential (11) we are limited by consideration of the first type of solution only.

It is visible from (12) - (14), that for the given depth of the quantum well and the photon energy (the difference of levels energies: $h\nu = E_3 - E_2 = E_2 - E_1$), the parameters of potential $U$, $L$, and the first energy level $E_1$ can be considered as functions of barrier thickness. The graphs given in figure 4 are calculated for the photon energy $h\nu = 117.16\,meV$ that corresponds to the wavelength $10.6\,\mu m$ [14]. In figures 4a and 4b the dependencies of $E_1$, $U$, $L$ from $d$ are shown taking into account the effective mass changes. In figures 4e and 4f similar dependencies are shown but without taking into account effective changes ($m = 0.067 m_e$). It is visible that the levels can only be equidistant for values of $d$ more than $4 nm$. The graphs show the increase of $d$ leads to the increase $E_1$, $U$ while dependence well of width $L$ from $d$ is non-monotonic function. It is important to note that at any value of $d$ taking into account the effective mass changes reduces the well width smaller to then the well width determined without the taking into account effective mass changes.

In figures 4c, 4d and 4g, 4h the dependencies of dipole matrix elements $\mu_{12}, \mu_{23}, \mu_{13}$, and their product $\mu_{12}\mu_{23}\mu_{13}$ from the barrier width $d$ as shown both with and without the effective mass. Note that at any value $d$ $\mu_{31} < \mu_{12} < \mu_{23}$. As it is seen from figures 4c and 4g the increase of $d$ decreases the transition probability between the first and second, and the second and third levels, while the electron transition probability from the third level to the first one increases. It should be mentioned that taking into account effective mass changes results in an increase of value of the SHG coefficient, in comparison with the case when the effective mass was taken constant. The maximum of generation is at $d = 5.53$.

In figure 5 the potential providing the maximal SHG for the photon wave length $10.6 \mu m$ is shown. As one can see from the figure the generation is maximal at $L \approx 10.3 nm$ and $d \approx 5.5 nm$. In the papers [14, 18] for experimental observation of SHG for wavelength $10.6 \mu m$ the heterostructure had been chosen with parameters $L \approx 10.5 nm$ and $d \approx 4.5 nm$ which were calculated taking into account the effect effective mass changes. Our calculations show that these values of parameters do not provide the equidistance of $E_2 - E_1$ and $E_3 - E_2$ (see figure 4f), and maximality the SHG coefficient also (see figure 4h). So, in particular, at $d \approx 4.5 nm$ the required equidistance takes place at $L \approx 10.86 nm$. If for experiment one use the heterosystem with values of parameters suggested in the present work the SHG well increases twice.

In figure 6 we present the table in which the maximum of the SHG coefficient and the potential parameters are given for different values of radiation wavelengths. As it can be seen the increase of radiation frequency decreases the coefficient maximum value.

## CONCLUSION

In this paper the problem of determination of the maximum of SHG for the potential well containing inside a rectangular barrier is considered. Consideration was carried out both for the case of infinite deep well and for the case of finite depth well. It is shown that the problem of determination of ensemble of structures with equidistant first three levels has two types of solutions. For the first type the second and third levels are located in underbarrier region and for the second type of solutions above the barrier the third level is located only. It is also shown that generation corresponding to the second type of solution always is less than generation for the second type.

The problem of finding the maximum of SHG coefficient for the finite depth well both with and without taking into account the effective mass changes is solved precisely. It is shown that the structural and composite parameters of the structural chosen earlier for experimental observation of the SHG can be essentially updated [18].

## Acknowledgement

We would like to thank prof. S.G.Petrosyan for useful discussion.

REFERENCES

1. **C.Weisbuch, B.Vinter**. Quantized Semiconductor Structures: Physics and Applications. Boston, Academic Press, 1991
2. **W.Trzeciakowski, B.D.McCombe**. Appl. Phys. Lett., **55**, 891 (1989).
3. **A.Lorke, M.Merkt, F. Malcher, G.Weimann, W. Schlapp.** Phys. Rev. B, **42**, 1321 (1990).
4. **M.Seto, M.Helm, Z.Moussa, P.Baucaud, F.H.Julien, J.M.Lourtioz, J.F.Nutzel, G.Abstreiter**. Appl. Phys.Lett., **65**, 2969 (1994).
5. **J.Khurgin.** Appl. Phys.Lett., **51**, 2100 (1987); Phys. Rev. B, **38**, 4056 (1988).
6. **E.Rosencher, P.Bois, J.Nagle, E.Costard, S.Delaitre.** Appl. Phys. Lett., **55**, 1597 (1989).
7. **E.Rosencher.** J.Appl. Phys., **73**, 1909 (1993).
8. **H.Kuwatsuka, Ishikawa.** Phys. Rev. B, **50**, 5323 (1994).
9. **L.Tsang, E.Ann, S.L.Chuang.** Appl. Phys. Lett., **52**, 697 (1988).
10. **T.Park, G.Gumbs, Y.C.Chen.** J.Appl. Phys., **86**, 1467 (1999).
11. **S.Tomic, V.Milanovic, Z.Ikonic.** Phys. Rev. B, **56**, 1033 (1997); J.Phys.: Condens. Matter, **10**, 6523 (1998).
12. **G.Goldoni, F.Rossi.** Optics Lett., **25**, 1025 (2000).
13. **E.Rosencher, P.Bois.** Phys. Rev. B, **44**, 11315 (1991).
14. **E.Rosencher, P.Bois.** *Intersubband Transitions in Quantum Wells*. Edited by E.Rosencher et al., New York, Plenum Press, 1992.
15. **D.M.Sedrakian, A.Zh.Khachatrian.** J. Contemp. Phys. Acad. Sci. of Armenia, **36**, 62 (2001).
16. **E.T.Yu et al.** Solid State Phys., **46**, 2 (1992).
17. **D.F.Nelson, R.C.Miller, D.A.Kleinman.** Phys.Rev. B, **35**, 7770 (1987).
18. **Ph.Boucaud, F.H.Julien, D.D.Yang, J-M.Lourtioz, E.Rosencher, Ph.Bois,J.Nagle.** Appl. Phys. Lett., **57**, 215 (1990).

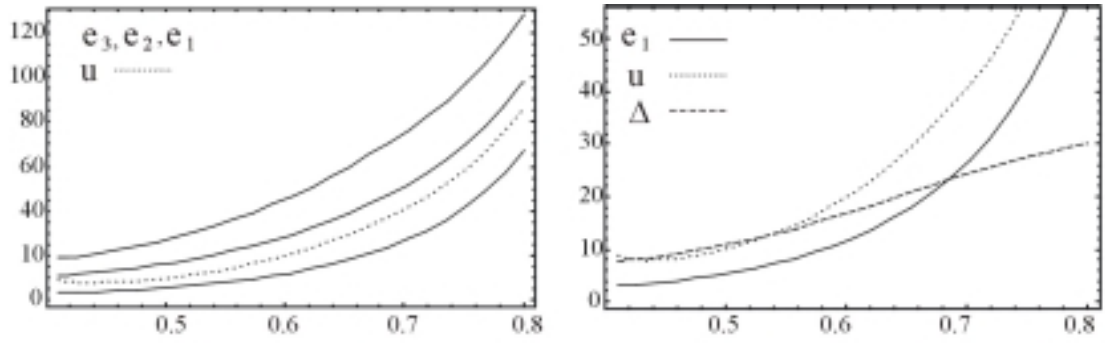

Figure 1. Dependencies of $e_1$, $e_2$, $e_3$, $u$ and $\Delta$ on the variable $x$ for the first type of solution.

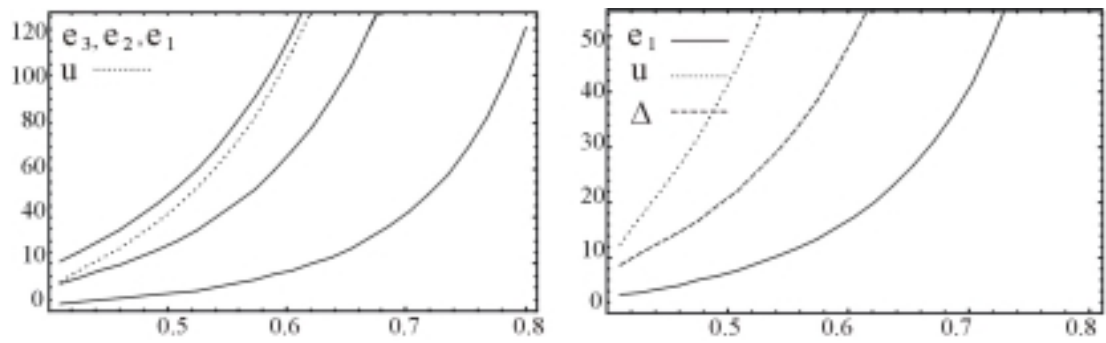

Figure 2. Dependencies of $e_1$, $e_2$, $e_3$, $u$ and $\Delta$ on the variable $x$ from the second type of solution.

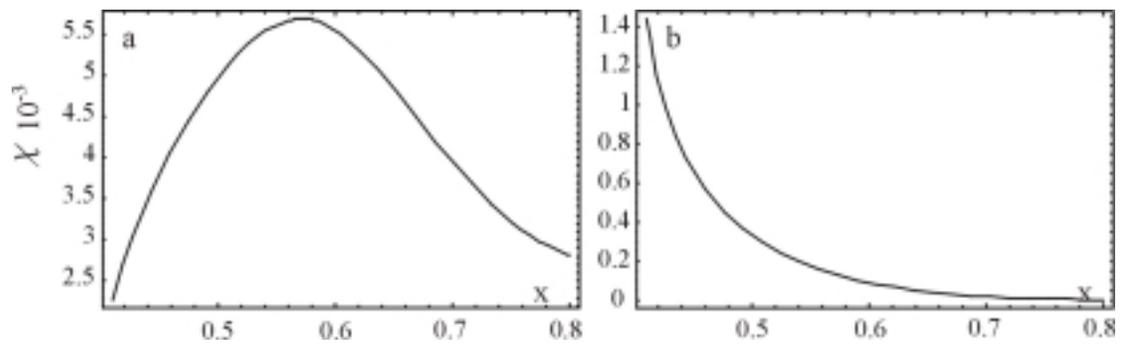

Figure 3. Dependence of $\chi$ on $x$ for the first (a) and the second (в) types of solution.

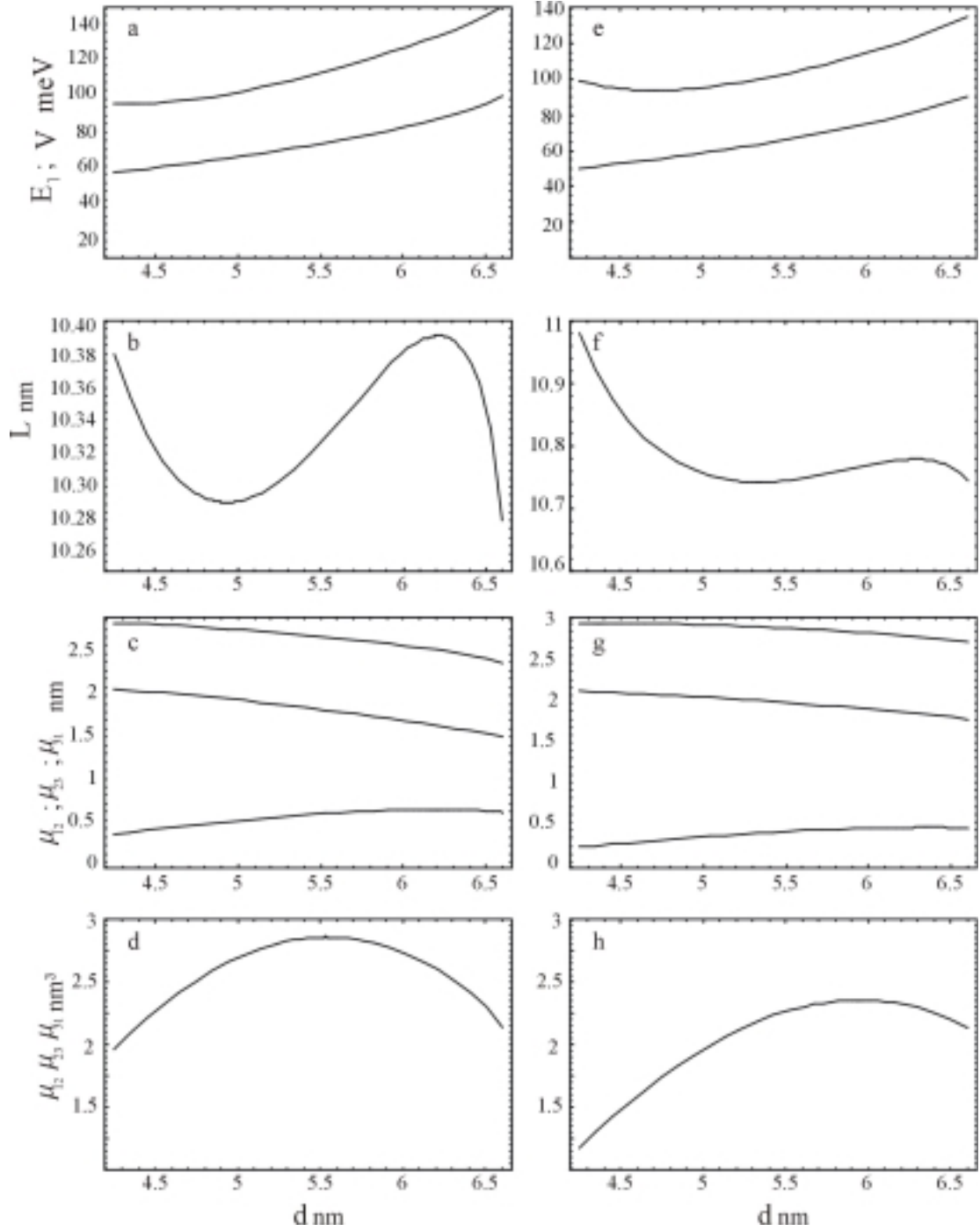

Figure 4. Dependencies of $E_1$; $U$; $L$; $\mu_{12}$; $\mu_{23}$; $\mu_{13}$; $\mu_{12}\mu_{23}\mu_{13}$ on the width of the potential barrier $d$ with and without taking into account the changes effective mass (see a), b), c), d) and e), f), g), h), correspondingly).

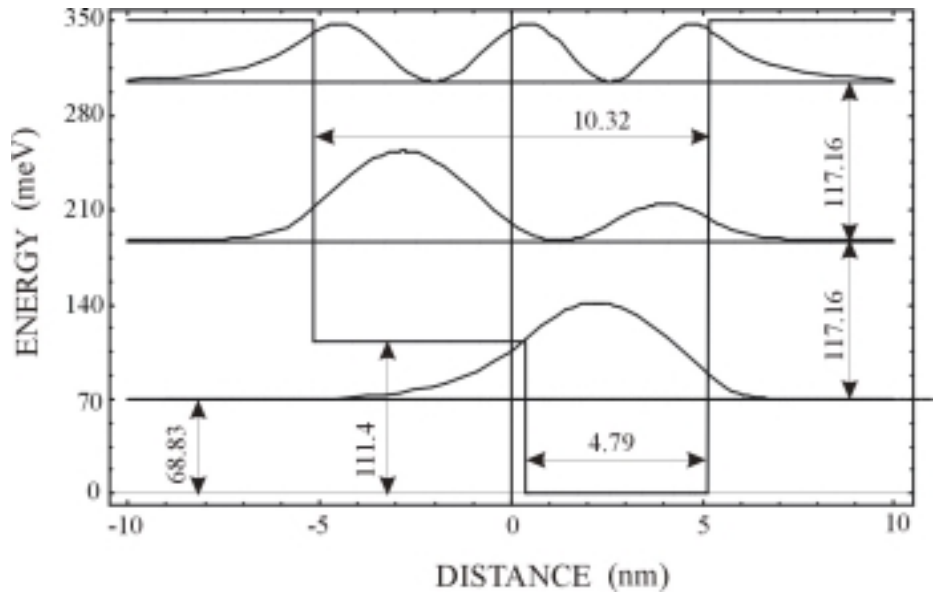

Figure 5. The quantum well providing the maximal generation for second harmonic at the wavelength $10.6\,\mu m$, and the square of a module the wave function for the first three levels.

| λ μm | $\mu_{12}\mu_{23}\mu_{13}$ nm³ | d nm | L nm | U *meV* | $E_1$ *meV* |
|---|---|---|---|---|---|
| 11 | 3.034 | 5.734 | 10.69 | 108.494 | 66.1 |
| 10.9 | 2.991 | 5.686 | 10.6 | 109.31 | 66.8 |
| 10.8 | 2.947 | 5.634 | 10.51 | 110 | 67.44 |
| 10.7 | 2.902 | 5.582 | 10.42 | 110.728 | 68.13 |
| 10.6 | 2.857 | 5.529 | 10.33 | 111.433 | 68.83 |
| 10.5 | 2.809 | 5.471 | 10.23 | 112.019 | 69.48 |
| 10.4 | 2.760 | 5.411 | 10.14 | 112.572 | 70.14 |
| 10.3 | 2.708 | 5.350 | 10.04 | 113.054 | 70.8 |
| 10.2 | 2.653 | 5.282 | 9.93 | 113.338 | 71.38 |
| 10.1 | 2.594 | 5.211 | 9.83 | 113.539 | 71.97 |
| 10 | 2.530 | 5.123 | 9.72 | 113.177 | 72.23 |
| 9.9 | 2.460 | 5.029 | 9.61 | 112.537 | 72.38 |
| 9.8 | 2.381 | 4.927 | 9.50 | 111.583 | 72.39 |
| 9.7 | 2.293 | 4.817 | 9.39 | 110.298 | 72.26 |
| 9.6 | 2.192 | 4.688 | 9.27 | 108.288 | 71.68 |
| 9.5 | 2.079 | 4.540 | 9.16 | 105.9 | 70.91 |
| 9.4 | 1.952 | 4.401 | 9.05 | 103.159 | 69.935 |
| 9.3 | 1.810 | 4.400 | 8.88 | 104.066 | 72.751 |
| 9.2 | 1.65 | 4.070 | 8.85 | 96.751 | 67.26 |

Figure 6. the maximum of the SHG coefficient and the potential parameters are given for different values of radiation wavelengths.